\documentclass{JINST}

\title{Xenon Bubble Chambers for Direct Dark Matter Detection}

\author{C. Levy$^a$\thanks{Corresponding
author.}~, S. Fallon$^a$, J. Genovesi$^a$, D. Khaitan$^b$, K. Klimov$^d$, J. Mock$^{a,c}$ and M. Szydagis $^a$\\
\llap{$^a$}University at Albany, State University of New York,\\
  1400 Washington Ave, Albany, NY 12222, USA\\
\llap{$^b$} University of Rochester,\\
Rochester, New York 14627, USA\\
\llap{$^c$}Lawrence Berkeley National Laboratory,\\
1 Cyclotron Road, Berkeley CA 94720, USA\\
\llap{$^d$}Stony Brook University,\\
  Stony Brook, NY 11794, USA\\
  E-mail: \email{clevy@albany.edu}}

\abstract{The search for dark matter is one of today's most exciting
  fields. As bigger detectors are being built to increase their
  sensitivity, background reduction is an ever more challenging
  issue. To this end, a new type of dark matter detector is proposed, a xenon
bubble chamber, which would combine the strengths of liquid xenon
TPCs, namely event by event energy resolution,
with those of a bubble chamber, namely insensitivity to electronic
recoils. In addition, it would be the first time ever that a dark
matter detector is active on all three detection channels, ionization
and scintillation characteristic of xenon detectors, and heat through
bubble formation in superheated fluids. Preliminary simulations show
that, depending on threshold, a discrimination
of 99.99\% to 99.9999+\% can be achieved, which
is on par or better than many current experiments. A prototype is
being built at the University at
Albany, SUNY. The prototype is currently undergoing seals, thermal,
and compression testing.}

\usepackage{lineno}
\usepackage{graphicx}
\usepackage{subfig}
\usepackage{caption}

\begin{document}

\section{The Search for Dark Matter}

According to the $\Lambda CDM$ model of cosmology, dark matter is thought to make up at least a quarter of the universe,
while the 75\% remaining is made of dark energy (71\% ) and baryonic
matter (4\%) \cite{Planck2013}. One of the most popular candidates for a dark matter particle is
the WIMP, Weakly Interacting Massive Particle, which should have a 
very low, yet detectable, cross section with normal
matter. Many experiments have thus been devised
in order to discover this elusive particle. 

While direct detection experiments use different techniques, they all rely on
observing, in one way or another, the direct scatter of a WIMP with a
target nucleus, thus inducing a recoil of the nucleus. While gammas and electrons
interact with the electron cloud of the atom, a process referred to as
an electronic recoil (ER), WIMPs and
neutrons give nuclear recoils (NR) by either interacting coherently with all nucleons in the
nucleus (spin-independent interaction) or via spin coupling to the
spin of the nucleus (spin-dependent interaction) \cite{Lewin1996}.

Because the interaction cross-section between a WIMP and a target
nucleus is smaller than $4 \times 10^{-46}\; cm^2$ for example for a $33\; GeV/c^2$ mass WIMP
\cite{LUX2015}, such experiments can only be successful if they
have no other background, which means that nothing else but a WIMP can
interact in the detector and produce a distinguishable signal. Background control is thus the number one
priority of direct dark matter experiments. To avoid background, the detectors are usually located deep
underground, either in a mine or under a mountain, where they will be
shielded from cosmic rays and muon-induced neutrons. In addition, to
limit the background, these detectors are made of highly radiopure
materials \cite{XENON100}\cite{LUXbg}. However, eliminating backgrounds entirely is nearly
impossible and so being able to discriminate between nuclear and
electronic recoils is of crucial importance, particularly since the backgrounds consist almost entirely of electronic recoils from gammas and betas.

Amongst the plethora of experiments available, xenon TPCs (Time
Projection Chambers)  and bubble
chambers are maybe the most different detector types. A typical two phase xenon
detector can measure the deposited energy of the recoil but has a
discrimination of about 99.6-8\%, which leads to a $4-2\times10^{-3}$
misidentification probability \cite{LUX} \cite{XENON100}
\cite{ZEPLINIII}. 
A bubble chamber on the other hand has nearly 0\% chance of
misidentification (of the order of $10^{-10}$) but has no direct
means of measuring the event-by-event energy deposited in the recoil
\cite{PICO}, although the global energy can be statistically 
inferred via threshold setting and temperature and pressure sweeping.

Until now, these techniques have been used separately in different
experiments, but we are now proposing to combine the strengths of
these technologies into one, to build a xenon bubble chamber. 

\section{Bubble Chambers}

The bubble chamber concept was first invented by D.A Glaser over 50
years ago \cite{Glaser52}.
It consists of a vessel filled with gas that is compressed to the
liquid phase. By having a smooth and clean vessel, and thus preventing
nucleation sites, the pressure can then be slowly lowered
adiabatically below the vapor pressure without triggering a phase
transition. The fluid is still in liquid state, while being at a
pressure where it should be gaseous. The fluid is thus superheated. Consequently, bubble chambers only work for certain sets of temperature
and pressure depending on
the target material.
In addition, to avoid nucleation sites in the rough plumbing above the
vessel, which would then ruin the stability of the superheated liquid
in the vessel, a layer of buffer fluid sits on top of the liquid. It is this
buffer fluid that is compressed, and that in turn compresses the
liquid \cite{PICO}.

As an incident particle interacts with the fluid, it
can deposit enough energy, in the form of heat, to warm up the liquid
and trigger a phase transition from liquid to
gas, therefore creating a bubble. Once a bubble is formed, pressure
is quickly applied in order to crush the bubble by reverting the gas
back to its liquid state. 
The process can then be repeated, the fluid put back into a
superheated state in order for a bubble to reform should a particle
interact in the detector.

The bubbles that are formed in these recoils
can be measured by acoustic sensors, by photographing
the bubble or by recording the pressure rise using
pressure transducers \cite{PICO} \cite{COUPP}.

Bubble chambers therefore use only one energy deposition channel, heat, in
order to try and detect dark matter.  
They are threshold detectors, either a bubble forms or it doesn't. To
form a bubble, they require a threshold energy, a highly
localized energy deposition $E_C$ within a
critical radius $R_C$. If this critical energy is not deposited within
the associated
critical radius, a bubble will not form.

In addition, different particles have different stopping
powers. While minimum ionizing particles could in principle be high
energy enough to form a bubble, they deposit their energy over long
tracks rather than locally, actually preventing the bubble formation.
A bubble chamber is thus blind to particles
with low stopping power, as they have for example only 1 in $10^{10}$ or greater chance of depositing enough energy in a
small enough radius to trigger a phase transition.

This stopping power dependence on particle type is especially
important for gammas, and  bubble chambers can be tuned to be insensitive
to electronic recoils, a considerable advantage for low background
searches like dark matter. A bubble chamber can thus be made
sensitive only to nuclear recoils, which have a high stopping power.

Bubble chambers have thus far been very successful in the
spin-dependent proton sector where they hold the current best sensitivity
\cite{PICO}.

\section{Scintillation Detectors}

Scintillation detectors are one of today's technologies of choice in the search
for dark matter. The world leading experiments with the best
sensitivity to traditional WIMP masses between 7 and 1000 $GeV/c^2$ are both using liquid xenon as scintillator
\cite{LUX} \cite{XENON100} while other competitive experiments using
cryogenic crystals also collect scintillation light
\cite{Cresst}.

The most successful scintillator experiments are dual phase Time Projection
Chambers (TPCs). In this type of TPC, a particle interacts in the liquid target material producing
a primary scintillation signal (S1) that is detected by two PMT arrays at
the bottom and the top of the TPC. At
the same time as the scintillation, an ionization signal is
produced. The electrons from the ionization are drifted up to the gas
phase where they interact with the gas and emit a secondary scintillation signal (S2) also detected
by both PMT arrays. 
This type of liquid xenon detector therefore uses two detection channels,
scintillation and ionization, in order to try and detect dark
matter. While the combination of the light pattern on the PMT arrays and the drift
time of the electrons allows to reconstruct the position of each event in 3D, the
recording of both signals provide information about the recoil
energy, allowing for event by event energy reconstruction, since the amount of light scales with the recoil
energy. Recording both signals also allows for electronic recoil discrimination via the
ratio of S2 to S1 \cite{LUX} \cite{XENON100}.


\section{New Concept: Scintillating Bubble Chamber}

The novel idea is to combine the strengths of both bubble chamber and
scintillator into a new, more advanced dark matter detector which will
have both the energy reconstruction of a scintillator and
the electronic recoil rejection of a bubble chamber, as well as use
all three detection channels, scintillation, ionization and heat. 

\subsection{Choice of liquid}

For applications in scintillating bubble chambers, only liquid scintillators can be
used. Among these, organic scintillators as well as liquid noble gases
like xenon or argon can be used. 

A bubble chamber with a fluorinated organic scintillator would be advantageous for searching
for spin dependent WIMP-proton interactions \cite{PICO}, and would serve as a powerful complement to the spin
independent searches, particularly if a discovery is made.
However, the advantage of liquid nobles over organic scintillators is
that they can definitely be used to measure charge, allowing
measurement of heat, scintillation and charge simultaneously.

Amongst liquid nobles, there are several target materials to choose
from. Using argon in a bubble chamber for example would have
the advantage of using pulse shape discrimination at higher energies,
but many photons are needed for this, and so this feature could not be used at the lowest energies where the WIMP signal should be the largest. However, using the discrimination of
bubble chambers would allow one to still achieve very low thresholds,
which cannot be achieved in current argon detectors \cite{Darkside}.

However, xenon is perhaps the best choice, as using it in a bubble
chamber would allow an improvement in the electronic recoil
discrimination, eliminating the $^{85}Kr$ and neutrino electron
scattering backgrounds, two of the most challenging expected
backgrounds with which to deal for the
next generation of xenon experiments \cite{LZCDR}. 

\begin{figure}[!]
\centering
\begin{tabular}{ll}
\includegraphics[width=8cm]{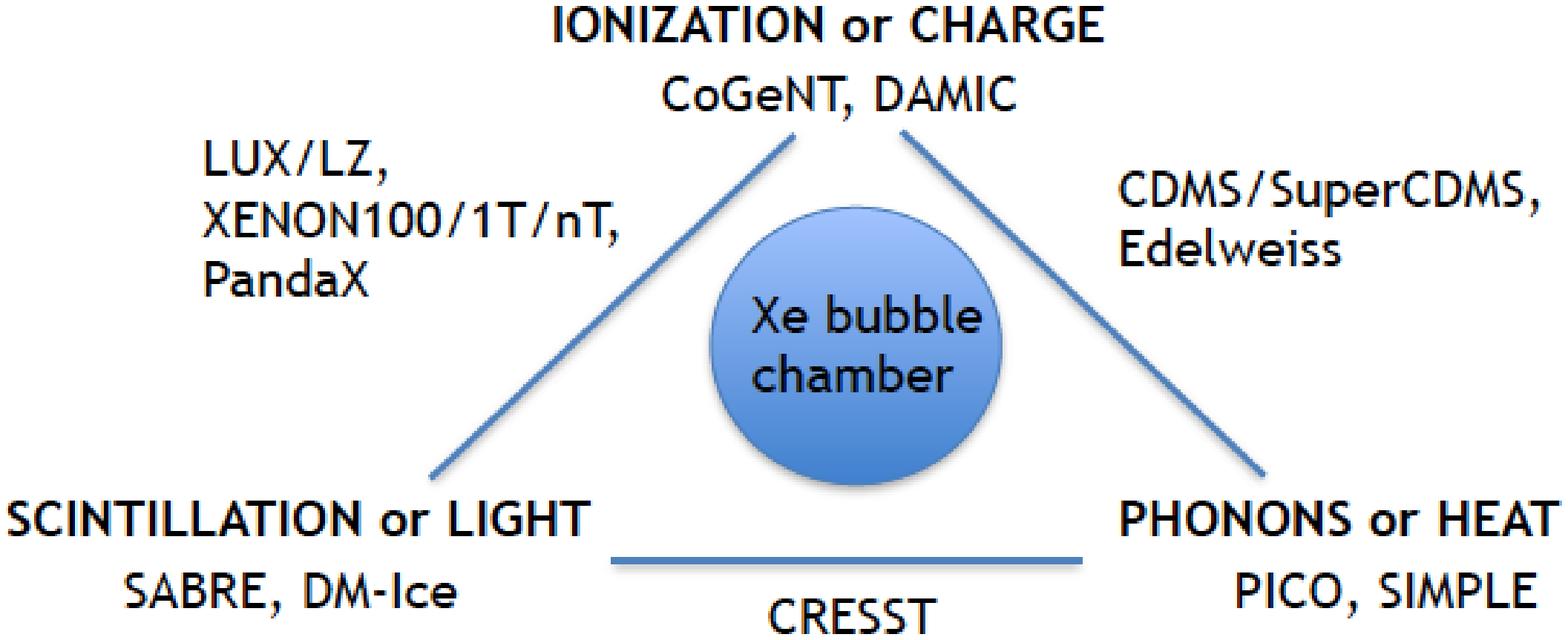} &

\includegraphics[width=7cm]{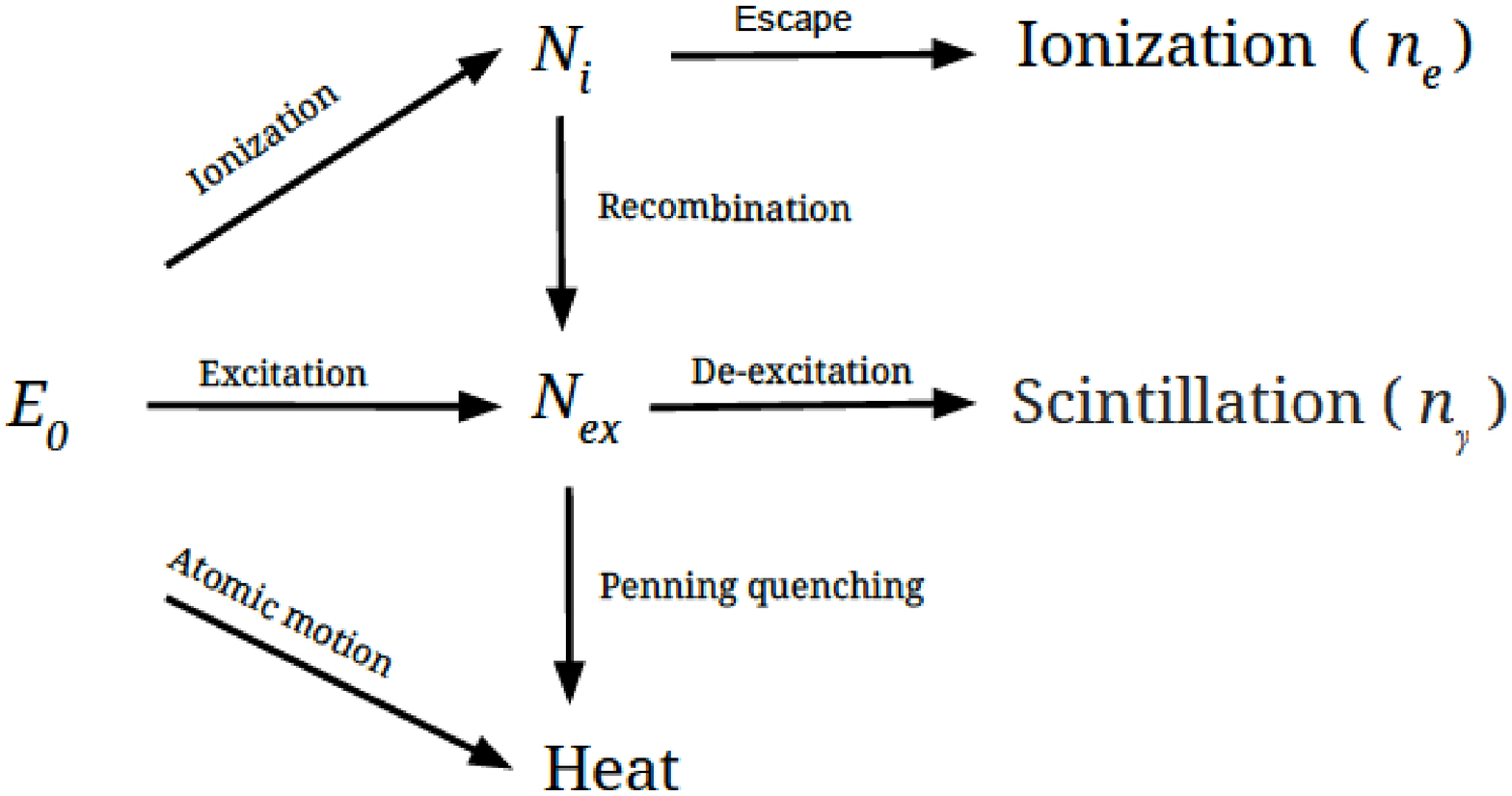}
\end{tabular}
\caption{Left: Detection channels of a bubble chamber. Right: Diagram
  of the energy deposition in liquid xenon into different detection
  channels \cite{Lenardo2014}.}
\label{fig:triangle}
\end{figure}

Moreover, as shown in Figure \ref{fig:triangle}, unlike all other
current dark matter experiments which only use one or two detection channels, a xenon bubble
chamber would be the first ever detector to attempt to use all
three detection channels, by recording heat as a bubble chamber,
scintillation as a liquid scintillator, and ionization by applying an
electric field to the vessel and drifting the charge through the
bubble, like is done in a TPC.

\subsection{1956 Xenon bubble chamber}

\begin{figure}[!]
\centering
\includegraphics[width=3cm]{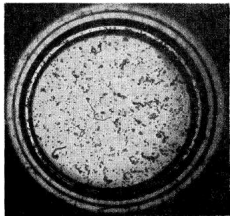}
\caption{Tracks from gamma rays in the original xenon bubble chamber
  doped with 2\% ethylene \cite{Glaser56}.}
\label{fig:bubblesGlaser}
\centering
\end{figure}

In 1956, Brown, Glaser and Perl attempted to use a xenon bubble
chamber \cite{Glaser56} to detect high energy gamma rays. At first the
experiment was a failure and no bubbles were observed, presumably due
to the energy being lost in scintillation rather than in heat. To
resolve this issue, the xenon was doped with ethylene to quench the scintillation signal. Bubbles from high energy gammas were then copiously observed as shown in Figure \ref{fig:bubblesGlaser}.

\noindent The logical consequence of these results is that, by not adding the
quenching material, and using a pure xenon target, a xenon bubble
chamber would indeed be blind to gamma rays while still generating
scintillation signal, which is exactly the desired goal for WIMP
detection. Moreover, because nuclear recoils already lose energy to
heat, nuclear recoils should make bubbles and a smaller-than-for-ER scintillation signal at the same time, thus providing an effective means of discrimination. Thus, this earlier measurement is one of the strongest motivations for this new WIMP detector, providing strong evidence that it should be successful.

\subsection{Advantages}

If both heat and scintillation can be observed in a xenon bubble
chamber, the scintillation signal can be
used for event by event energy reconstruction just like in TPCs, while
the heat signal can be used for electronic recoil
discrimination, just like in bubble chambers. 

Already, a two channel xenon bubble chamber would
combine the strengths of  bubble chambers and liquid xenon TPCs,
and thus outperform them in theory.
If a third detection channel, an S2 signal coming from
ionization, could be added to the scintillation and heat, a xenon
bubble chamber would be able to go lower in threshold while
maintaining its energy reconstruction capacity and blindness to ER, a
feat that no other dark matter experiment has thus far achieved.
To obtain an S2 signal, electrons would have to be drifted through the
bubble itself. This has already been attempted by \cite{Arazi2015} in non-superheated, stable
bubbles with promising first results \cite{Erdal2015}.

\begin{figure}[!]
\centering
\includegraphics[width=8cm]{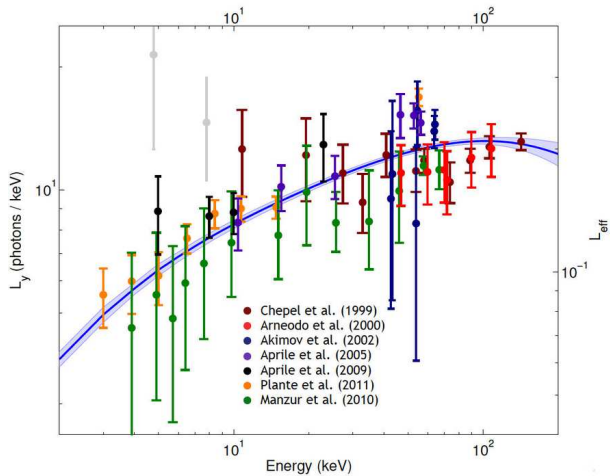}
\caption{Absolute light yield of liquid xenon (solid blue
  line) for NR with statistical error band using NEST which is a
  global fit to the world's data in all channels \cite{Lenardo2014}
  \cite{NEST} \cite{NEST2013}.}
\label{fig:Leff}
\centering
\end{figure}

In addition, a xenon bubble chamber has one more very important
advantage. Xenon detectors perform generally less adequately at low
energies due to reduced S1 and S2 signals from nuclear recoils as shown in
Figure \ref{fig:Leff} for example. This reduction comes from the efficiency of
nuclear recoils to generate more nuclear recoils. As a particle
interacts in the liquid, the recoiling nucleus will interact with
other nuclei which will in turn also recoil, instead of exciting or
ionizing atoms. Thus most of the energy
will be going to heat rather than to scintillation and is then lost for typical xenon
TPCs.

However, in a xenon bubble chamber, this energy would just go
from one channel to another; scintillation and ionization would be
reduced but the loss of energy would be caught via the heat channel. This means
that xenon bubble chambers should not suffer any efficiency loss at
low energies and could probe a low energy WIMP mass sector
that is unavailable to current experiments.

This also means that with a xenon bubble chamber, all the energy would
be accounted for and summing all three energy channels, light yield
from S1, charge yield from S2 and the bubble formation
probability from heat, would give total quanta independent of energy
or electric field. This would be an unprecedented calibration
opportunity to inform mainstream liquid xenon dark matter experiments.

\subsection{Prototype at University at Albany SUNY}

\begin{figure}[!]
\centering
\includegraphics[width=13cm]{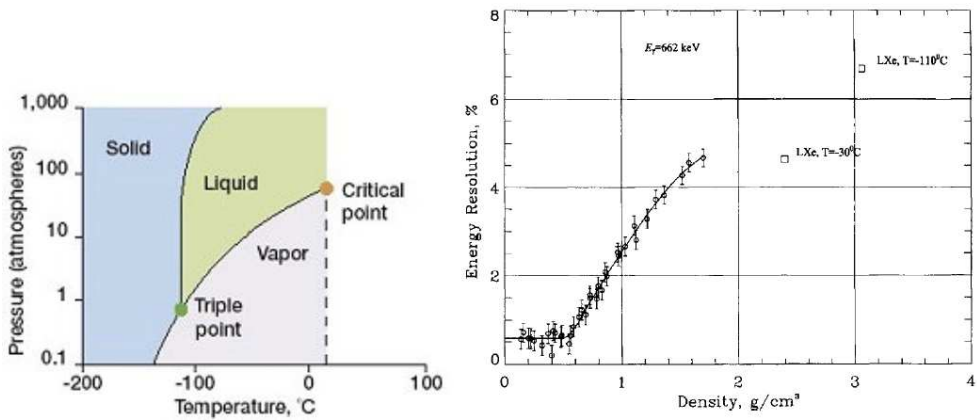}
\caption{Left: Phase diagram of xenon [Credit: NASA]. Right: Improvement of the energy resolution based on density for 662
keV gamma rays in liquid and gas xenon in the charge channel alone \cite{Ramsey97}.}
\label{fig:ERes}
\end{figure}

A prototype of a xenon bubble chamber is being developed at UAlbany.
Unlike xenon TPCs which operate at -100 $^\circ$C and typical bubble
chambers which operate at close to room temperatures and higher, this prototype will be
operated at -40 $^\circ$C, a temperature at which xenon can be 
superheated as shown in Figure
\ref{fig:ERes} left.

One possible impact of researching the response of liquid xenon
detectors at -40 $^\circ$C for xenon bubble chambers is that the
energy resolution and therefore S2/S1 discrimination can be very
different at this temperature than at -100 $^\circ$C as in typical
TPCs. Figure \ref{fig:ERes} (right) shows how the energy resolution improves at higher temperature for 662 keV gammas.


If this improvement in energy resolution for ER with increasing
temperature were to also apply at low energies, and for NR as well,
the S2/S1 bands for both NR and ER bands \cite{XENON100} \cite{LUX} would tighten and the
discrimination would consequently improve.  

Moreover, if energy resolution and electronic recoil
discrimination are found to improve at this higher temperature,
current dark matter experiments would profit from this knowledge while
future larger dark matter experiments
could be tuned, even slightly depending on design constraints, to try
and capitalize on this effect and improve their energy
resolution, and thus have a better chance of conclusively discovering dark matter.

In addition, xenon bubble chambers can help current bubble chamber
experiments such as PICO \cite{PICO}, which have an unknown background possibly
due to alphas straggling out of particulates, chemistry effects or
micro-droplets of buffer fluid. Adding a scintillation signal to these
experiments may help in identifying these backgrounds and by doing so,
achieve better results in the spin dependent sector. Indeed, a xenon
bubble chamber, having the energy information that typical bubble
chambers lack, will be able to identify alphas, whose energy on the
order of MeV is orders of magnitude larger than that of dark matter
recoils, by the S1 and S2 size and by pulse shape discrimination \cite{Dawson2005}.

To investigate these topics, and to prove the principle of a
scintillating xenon bubble chamber, a prototype chamber is being
developed at UAlbany. This chamber will hold between 10 and $\sim$ 200 g of xenon, and will
record bubble formation and scintillation simultaneously.

\begin{figure}[!]
\centering
\begin{tabular}{ll}
\includegraphics[width=8.5cm]{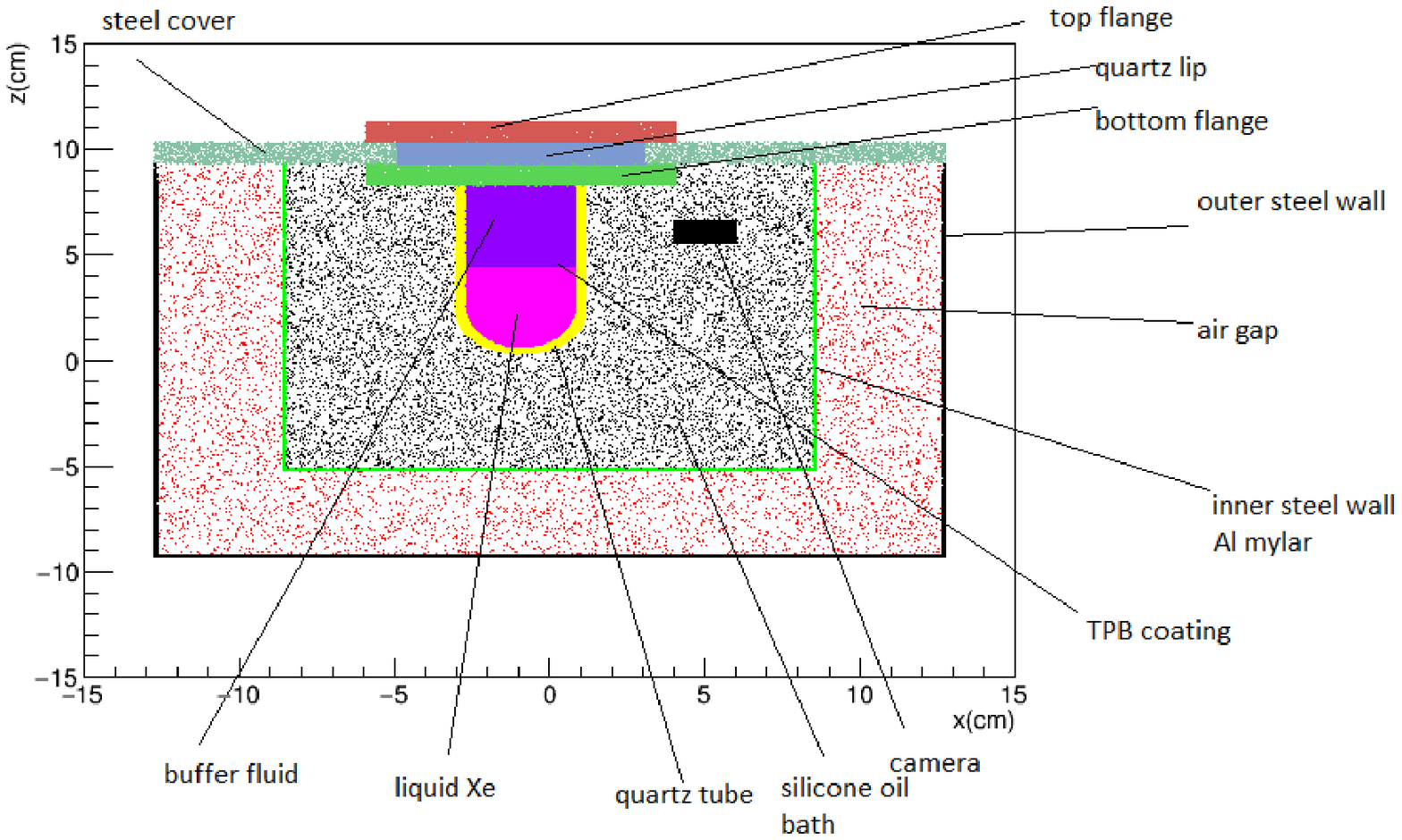} &

\includegraphics[width=6cm]{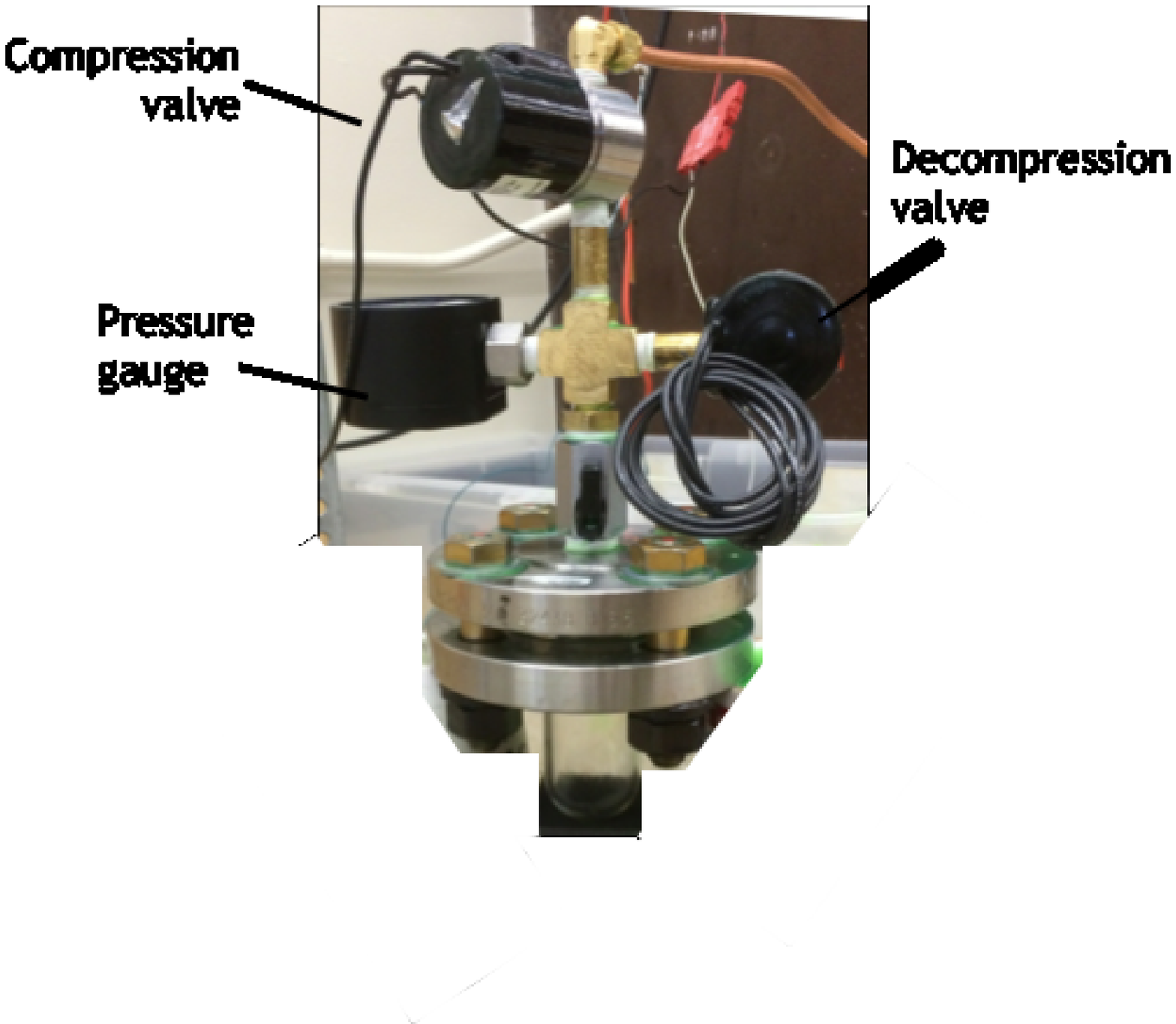}
\end{tabular}
\caption{The UAlbany bubble chamber. Left: GEANT4 simulation. Right:
  Current prototype for pressure/thermal cycling and
  seal tests with N2 gas. The decompression valve, linked to a gas recuperation system, will be used when tests with LXe will start.}
\label{fig:prototype}
\end{figure}

The UAlbany chamber itself, shown in Figure \ref{fig:prototype}, consists of a small quartz vessel filled with
liquid xenon with a buffer fluid resting on top to allow for
pressurization. The buffer fluid used is silicone oil and is also
the hydraulic and thermal bath fluid of choice. By having a thick
(3.25 mm) quartz tube capable of withstanding high pressures and the
same fluid as hydraulic, thermal and buffer, there is no
need for an external pressure vessel and bellows for pressure
equalization. The chamber will be operated at -40$^{circ}$C anywhere below
300 psia at maximum compression. The decompression pressure will
actually be varied to study different energy thresholds, for example as low as ~0.5
keV at -40${^\circ}$C \cite{Seitz1958}. The quartz vessel is
coated with TPB \cite{Gehman2013}, a transparent
wavelength shifter, in order to shift the light from VUV to
visible wavelengths. The bubble itself and potentially, the scintillation
light itself, will be recorded for the first time ever via a CCD
camera. Recording both bubble and scintillation light with
the same device will thus make the chamber setup more efficient. In
addition, a typical PMT with
high QE in blue/violet will be used to record the scintillation
light if the CCD fails to record it.
In order to reflect scintillation light into the PMT or CCD, the chiller
containing the thermal bath will be coated in aluminized mylar.

\begin{figure}[!]
\centering
\includegraphics[width=9cm]{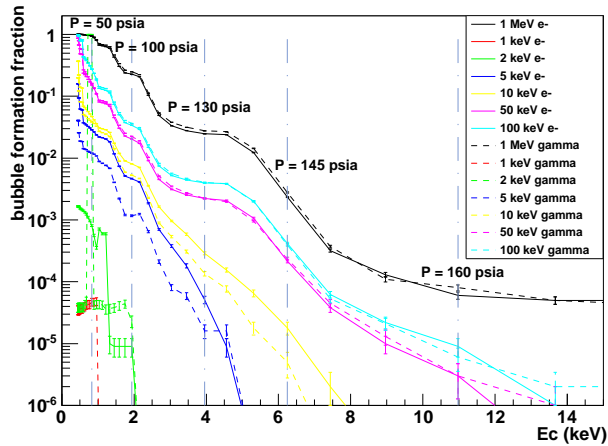}  

\caption{Simulations of the fraction of bubble formation dependency on
  critical energy $E_c$ at -40 $^\circ$C
  \cite{Seitz1958}. The vertical dashed lines represent different
  operating pressures. The energy axis approximates keVnr because
  virtually all NR above the critical energy will create bubbles as NR
  are more efficient at losing energy to heat than to S1 and S2. The y axis approximates the recoil discrimination but is
  a conservative estimate as it assumes all energy goes into heat and
  neglects the energy lost in scintillation and ionization.}
\label{fig:mysim}
\end{figure}

This prototype should address two questions. First, it will allow for a
study of what the gamma discrimination is as a function of recoil
threshold.
This has already been studied in simulations using the Seitz model \cite{Seitz1958}, NEST \cite{NEST} , GEANT4 \cite{GEANT4} and
PENELOPE \cite{PENELOPE}. Figure \ref{fig:mysim} shows the fraction of bubble formation as a
function of critical energy $E_C$ for electronic recoils due to 1 MeV, 100,
50, 10, 5, 2, and 1 keV  
electrons and gammas in liquid xenon. The steps correspond to
the electronic shells of xenon. The plateau at higher energies is due to the
resolution limit of the simulation software. This shows that xenon
bubble chambers could in principle reach 99.9999\% or better
discrimination, so up to a
factor 4000 better than standard liquid xenon detectors, depending on threshold.

The second question to answer will be whether
or not the usage of a buffer fluid will quench the
scintillation. Liquid noble detectors are notoriously ultrapure
systems to prevent the scintillation light from being absorbed by the
impurities in the target material
before reaching the PMTs. The buffer fluid thus needs to be completely
immiscible and separate perfectly
from the liquid xenon, as even the smallest amount of contamination
may jeopardize the light collection. New techniques might need to be devised in
order to prevent this from happening.

For now, the UAlbany chamber is undergoing pressure and thermal
cycling tests as well as sealing tests. In the meantime, a
collaborating group at Rensselaer Polytechnic Institute is
investigating new electric field designs to apply to the UAlbany chamber in order to drift electrons
through a bubble and acquire an ionization signal in addition to the
scintillation and heat signals. A second decompression line will
eventually be added in order to
decompress below 1 atm in order to reach low thresholds.

\section{Conclusion}

A new type of dark matter detector is proposed, a scintillating xenon
bubble chamber which would have the potential for high discrimination
against electronic recoils at low thresholds like a typical bubble
chamber while also recording all the energy information like a typical
xenon TPC. 
Preliminary simulations give a conservative estimate on the predicted
discrimination from xenon bubble chambers of 99.99\% to 99.9999+\%. 
In addition, scintillating xenon bubble chambers may be the first detectors
to ever use all three discovery channels, heat, scintillation and
ionization, making it the most versatile and powerful dark matter
detector. 
In theory, provided a change in temperature and pressure, such a detector could potentially be expanded beyond the
initial use with xenon, and several scintillating bubble chambers with different target
materials could be used to help diagnose problems and improve other types of dark matter
experiments, as well as confirm the expected A$^2$ dependence of any
potential WIMP signal. A first prototype at University at Albany SUNY is currently under construction.

\acknowledgments
The authors would like to thank Prof. Mani Tripathi, Dr. Scott
Stephenson and Dustin Stolp of the UC Davis group for the TPB coating
of the quartz vessel, as well as Prof. Eric Dahl, Dr. Jianjie Zhang
and Dan Baxter of Fermilab and Northwestern University for fruitful talks, advice and
data sharing.

\end{document}